\documentclass[final,5p,times,twocolumn]{elsarticle}
\usepackage{latexsym,graphicx,amssymb,amsmath,mathrsfs}
\usepackage[usenames]{color}
\usepackage[colorlinks=true, pdfstartview=FitV, linkcolor=red, citecolor=blue, urlcolor=blue]{hyperref}
\usepackage{setspace,bm}
\usepackage[caption=false]{subfig}
\usepackage{amsmath}
\usepackage{latexsym}

\usepackage[caption=false]{subfig}
\usepackage{amscd}
\graphicspath{{./Figs/}}

\newcommand\beq{\begin{equation}}
\newcommand\eeq{\end{equation}}

\usepackage[normalem]{ulem}  % \sout{old text} for strikeout

\newcommand{\comment}[1]{}

\renewcommand\sout{\bgroup \color{red} \ULdepth=-.5ex \ULset}
\usepackage{amssymb}
\usepackage{amsmath}

\journal{Physics Letters B}

\begin{document}

\begin{frontmatter}



\title{Lefschetz thimbles in fermionic effective models with
       repulsive vector-field\tnoteref{Report}}
\tnotetext[Report]{Report number:KUNS-2679, YITP-17-50}

\author[Kyoto]{Yuto Mori}
\author[YITP]{Kouji Kashiwa}
\author[YITP]{Akira Ohnishi}

\address[Kyoto]{Department of Physics, Faculty of Science, Kyoto
University, Kyoto 606-8502, Japan}
\address[YITP]{Yukawa Institute for Theoretical Physics, Kyoto University,
Kyoto 606-8502, Japan}

\begin{abstract}
We discuss two problems in complexified
 auxiliary fields in fermionic effective models,
 the auxiliary sign problem associated with the repulsive vector-field
 and the choice of the cut for the scalar field appearing from the
 logarithmic function.
In the fermionic effective models with attractive scalar
 and repulsive vector-type interaction,
 the auxiliary scalar and vector fields appear in the path integral
 after the bosonization of fermion bilinears.
When we make the path integral well-defined
 by the Wick rotation of the vector field,
 the oscillating Boltzmann weight appears in the partition function.
This ``auxiliary'' sign problem can be solved
 by using the Lefschetz-thimble path-integral method,
 where the integration path is constructed in the complex plane.
Another serious
 obstacle in the numerical construction of Lefschetz thimbles
 is caused by singular points and cuts
 induced by multivalued functions of the complexified scalar field
 in the momentum integration.
We propose a new prescription which
 fixes gradient flow trajectories on the same Riemann sheet in the flow
 evolution by performing the momentum integration in the complex domain.
\end{abstract}

\begin{keyword}
Sign problem, Complex action


\end{keyword}

\end{frontmatter}

% Introduction
%%%%%%%%%%%%%%%%%%%%%%%%%%%%%%%%%%%%%%%%%%%%%%%%%%%%%%%%%%%%%%%%%%%%%%%%

\section{Introduction}
\label{Sec:Intro}

{\em General introduction:}
The sign problem appearing in the path integral is a serious
obstacle to perform precise nonperturbative computations
in various quantum systems:
The Boltzmann weight in the partition function oscillates and then it
induces the serious cancellation to the numerical integration process.
Particularly, the sign problem attracts much more attention recently in
the lattice simulation of Quantum Chromodynamics (QCD) at finite
density.
It is caused by the combination of the gluon field ($A_\mu$)
and the real quark chemical potential ($\mu$) in the
fermion determinant of the Boltzmann weight;
see Ref.~\cite{deForcrand:2010ys} for a review.

{\em Approaches to sign problem:}
Several methods have been proposed to circumvent the sign problem such as
the multi-parameter reweighting method~\cite{Fodor:2001au,Fodor:2001pe},
Taylor expansion method~\cite{Allton:2002zi,Allton:2005gk,Gavai:2008zr},
the imaginary chemical potential
approach~\cite{Lombardo:1999cz,deForcrand:2002hgr,D'Elia:2002gd},
the canonical
approach~\cite{Dagotto:1989fw,Alford:1998sd,Miller:1986cs,Hasenfratz:1991ax}
and so on.
These methods can be applied to nonzero $\mu$,
but we can not obtain reliable results in the large $\mu$ region

{\em Complex Langevin and Lefschetz thimble:}
Recently, two approaches for the lattice simulation
have been attracting much more attention;
the complex Langevin method and the Lefschetz-thimble path-integral method.
The complex Langevin method is based on the stochastic
quantization~\cite{Parisi:1984cs,Klauder:1983sp,Aarts:2009uq,Fodor:2015doa},
it does not use the standard Monte-Carlo sampling,
and thus it seems to be free from the sign problem.
However, this method sometimes provides a wrong
answer when there are singularities of the drift term in the
Langevin-time evolution~\cite{Aarts:2009uq,Nishimura:2015pba}.
By comparison, the Lefschetz-thimble path-integral
method~\cite{Witten:2010cx,Cristoforetti:2012su,Fujii:2013sra}
is based on the Picard-Lefschetz theory for the complexified
space in variables of integral~\cite{pham1983vanishing} and thus it is
still in the framework of the usual path-integral formulation.
With this method, we modify the integration path from the original
one to new one on which the complex phase is constant and thus the cancellation
is suppressed.
Thus, we can soften the difficulty of the sign problem.

{\em Sign problem in effective models:}
In effective models of the fundamental theory, one can sometimes avoid
the sign problem because of the simplification of the Boltzmann weight.
For example, in the standard Nambu--Jona-Lasinio (NJL) model, one of
the low-energy effective models of QCD, one can avoid the sign
problem at finite $\mu$ due to the simplified
fermion determinant.
The Dirac operator of the NJL model has the $C\gamma_5$ hermiticity 
and its determinant is real,
$\mathrm{det}\{{\cal D}_\mathrm{NJL}(\mu)\}
 = [\mathrm{det}\{{\cal D}_\mathrm{NJL}(\mu)\}]^*$,
where $C$ is the charge conjugation matrix.
However, the sign problem comes back, when the repulsive
vector-current interaction is included in the NJL model
and the auxiliary vector field is Wick
rotated to make the path integral well-defined.
This type of the sign problem which we call the {\it auxiliary sign
problem} in this paper has not been discussed in the
four-dimensional space-time, previously.
Understanding the auxiliary sign problem is very important because
it appears not only in the NJL model but also in
several calculations
which include repulsive interactions between fermions:
For example, the relativistic mean-field (RMF) models
with vector-meson ($\omega$) field in nuclear physics
are successful with the prescription that takes the saddle point value
for the temporal component of the vector field,
while they encounter the same auxiliary sign problem as the NJL model
when fluctuations are considered.
The shell model Monte-Carlo method also has the problem;
the Hubbard-Stratonovich transformation of repulsive two-body interactions
leads to the auxiliary sign problem, then an analytic continuation
from the attractive region is needed.
See Ref.~\cite{Koonin:1996xj} for a review.

In this paper, we try to apply the Lefschetz-thimble method to the
auxiliary sign problem of fermionic models.
To solve or assuage the auxiliary sign problem, it is
natural to apply
the Lefschetz-thimble path-integral method as in the lattice calculation.
Very few attempts of the Lefschetz-thimble path-integral method for the
auxiliary sign problem have been
done~\cite{Fujii:2015bua,Fujii:2015vha,Alexandru:2016ejd} in the
Thirring model~\cite{Thirring:1958in}.
We show how this method resolve the auxiliary sign problem and how it
makes the path integral well-defined.
In addition, we discuss the difficulty induced by singular points and
cuts in the complex plane of variables of integration.
To show the solving procedure explicitly, we employ the NJL model with the
vector-current interaction which is transformed into the repulsive
vector-field after the bosonization of fermion bilinears.
The NJL model is widely used
not only in hadron physics but also in the physics beyond the standard
model~\cite{Hasenfratz:1991it,Rantaharju:2016jxy} and
the dark matter phenomenology~\cite{Kubo:2014ida,Channuie:2016iyy}.
To obtain the analytic form of the effective potential,
we use the homogeneous auxiliary-field ansatz which can be acceptable
if the inhomogeneous phases~\cite{Nakano:2004cd,Nickel:2009ke} do not appear.

This paper is organized as follows.
Section~\ref{Sec:GF} shows details of the Lefschetz thimble method.
In Sec. \ref{Sec:EP},
we explain the formalism of the NJL model.
In Secs.~\ref{Sec:NR} and \ref{Sec:omega}, we discuss singularities induced
by multiple-valued functions in the momentum integration and show the
prescription for it in systems which have only the auxiliary scalar or
vector fileds, respectively.
Section~\ref{Sec:Summary} is devoted to summary.

\section{Gradient flows and Lefschetz thimbles}
\label{Sec:GF}
The Lefschetz thimbles can be obtained by solving gradient flows;
\begin{align}
 \frac{d z_i}{d t}
 &=  \overline{ \Bigl( \frac{\partial \Gamma [z]}
                            {\partial z_i} \Bigr)},~~~
\frac{d z_i}{d t}
  = -\overline{ \Bigl( \frac{\partial \Gamma [z]}
                            {\partial z_i} \Bigr)},~~~
\label{Eq:gf}
\end{align}
where $z_i$ means the complexified variables of integration,
$x_i' \to z_i \in \mathbb{C}$ and $t \in \mathbb{R}$ is the fictitious
time.
The fixed point of gradient flows are obtained from
\begin{align}
\frac{\partial \Gamma[z]}{\partial z_i} &= 0.
\end{align}
The first and second gradient flows in Eq.~(\ref{Eq:gf})
provides the downward and upward flows respecting
the Morse function, $h=-\mathrm{Re} (\Gamma[z])$.
Downward flows starting from fixed points describe new integration
paths (${\cal J}$) which
are so called the Lefschetz thimbles if corresponding upward-flow
trajectories (${\cal K}$) go across the original integration path.

On the Lefschetz thimbles, we can prove
\begin{align}
 \frac{d}{d t} \mathrm{Im} (\Gamma [z]) = 0,~~~~
 \frac{d}{d t} \mathrm{Re} (\Gamma [z]) > 0.
\label{Eq:gfc}
\end{align}
Thus, the sign problem seems to be resolved because
$\mathrm{Im}(\Gamma[z])$
is constant on the Lefschetz thimble and thus oscillation vanishes.
However,
there are remnants of the original sign problem.
One is the global sign problem which arises when multi-thimbles become
relevant to the integral.
The grand-canonical partition function can be decomposed into the
summation in terms of Lefschetz thimbles as
\begin{align}
 {\cal Z}
 &= \int_{C_\mathbb{R}} d^n x~e^{-\Gamma[x]}
  = \sum_{\tau} n_\tau \int_{\mathfrak{J}_\tau} d^nz~e^{-\Gamma[z]},
\end{align}
where $n_\tau$ is the crossing number of the upward flow with the
original integration-path and $\tau$
characterizes each Lefschetz thimble, ${\cal J}_\tau$.
Thus, there may be the cancellation in the numerical integration if each
relevant thimble has a different constant value of $\mathrm{Im}(\Gamma[z])$.
At present, there is no way to exactly solve the global sign problem in
the lattice simulation,
but it does not matter in the following discussions
and thus we leave it as a future work.
One of the promising approach to avoid the global sign problem has been
proposed in Ref.~\cite{Alexandru:2016ejd,Fukuma:2017fjq} by modifying the original
integration-path contour by using the gradient flow.
The other is the residual sign
problem which comes from the Jacobian of the new integration-path
contour.
The residual sign problem seems to be controlled by the phase reweighting
method at present; see Ref.~\cite{Fujii:2013sra}.

\section{Effective potential in the NJL model}
\label{Sec:EP}

The Euclidean action
of the two-flavor three-color NJL model
is expressed as
\begin{align}
\Gamma_\mathrm{NJL}
= \int d^4x_E &\left[{\bar q} (-i\gamma_\mu \partial_\mu + m_0 -\mu\gamma_0) q
- G [({\bar q}q)^2 + ({\bar q} i \gamma_5 \vec{\tau} q)^2]
\right.
\nonumber\\
&\left. 
 + G_\mathrm{v}({\bar q} \gamma_0 q)^2
 - G_\mathrm{v}({\bar q} \gamma_i q)^2
\right],
 \label{Eq:NJL}
\end{align}
where $q$ denotes the quark field, $m_0$ is the current
quark mass, and $\mu=1, \cdots, 4$ with $x_4=\tau=it$ and
$\gamma_4=i\gamma_0$.
We consider the case where $G>0$ and $G_\mathrm{v}>0$.
The last two terms are
nothing but the vector-current interaction which leads to
the repulsive vector-field after the bosonization of
the effective action.
Coupling constants $G$ and $G_\mathrm{v}$ are related with each other via
the Fierz transformation of the one-gluon exchange interaction; see appendix of
Ref.~\cite{Kashiwa:2011td} as an example.

The auxiliary sign problem appears as a consequence of
defining the path integral of the auxiliary vector field
by using the Wick rotation.
The grand canonical partition function after the bosonization of quark
bilinears by using the Hubbard-Stratonovich transformation
is formally written as
\begin{align}
 {\cal Z}
 &= \int {\cal D}q {\cal D}\bar{q} ~e^{ - \Gamma_\mathrm{NJL} [q,\bar{q}]}
  = \int_{C_\mathbb{R}}
         {\cal D} \sigma{\cal D} {\vec \pi} {\cal D} \omega_\mu
         ~e^{ - \Gamma [\sigma,{\vec \pi},\omega_\mu]},
\\
\Gamma
&= -\log \det D + \int d^4x_E \left[
G (\sigma^2(x)+{\vec\pi}^2(x))
+G_\mathrm{v} \omega_\mu^2(x)
\right]\ ,
\\
D&=-i\gamma_\mu\partial_\mu + M -\gamma_0\mu'
-2iG\gamma_5{\vec\pi}\cdot{\vec\tau}
+2G_\mathrm{v}\gamma_i\omega_i\ ,
\end{align}
where
$M(x)=m_0+2G\sigma(x)$, $\mu'=\mu-2iG_\mathrm{v}\omega_4(x)$,
$C_\mathbb{R}$ means the integration path in the real variables.
The variables of integration, $\sigma$, ${\vec \pi}$ and $\omega_\mu$
with $\mu = 1, \cdots, 4$, are the scalar, pseudo-scalar and vector
mesonic fields after using the Wick rotation for the $\omega_0$ field,
respectively.

In the homogeneous auxiliary-field ansatz,
we can simplify $\Gamma$ as $\Gamma = \beta V {\cal V}$ where ${\cal V}$
corresponds to the effective potential.
In the concrete calculation, we should consider three different-type
variables of integration, $X=(\sigma,{\vec \pi},\omega_\mu)$,
but we here only consider the limited set,
$X'=(\sigma,\omega_4)$,
which provide the minimal set to discuss the auxiliary sign
problem.
The analytic form of ${\cal V}$ becomes
\begin{align}
 {\cal V}
 &= - 2 N_\mathrm{f} N_\mathrm{c} \int \frac{d^3 p}{(2\pi)^3}
      \Bigl[ E_p + T ( \ln f^- + \ln f^+ ) \Bigr]
\nonumber\\
 &  \hspace{4mm}
    + G \sigma^2 + G_v \omega_4^2,
\label{Eq:EP-NJL}
\end{align}
where $N_\mathrm{f}=2$, $N_\mathrm{c}=3$ and
$ f^\mp=1+e^{-\beta (E_p \mp \mu')} $
with $E_p = \sqrt{{\bf p}^2 + M^2}$.
The constituent quark mass ($M$) and the effective real chemical
potential ($\mu'$) becomes
$M = m_0 + 2G \sigma$ and $ \mu' = \mu - 2 i G_\mathrm{v} \omega_4$.
The expectation values of $\sigma$ and $\omega_4$ are
$\langle \sigma \rangle = \langle {\bar q} q \rangle$ and
$\langle \omega_4 \rangle = - i\langle \omega_0 \rangle
                          = - i \langle q^\dag q \rangle$.
With the Wick rotation
for the $\omega_0$ field,
$\mu'$ takes complex values and then the effective
action becomes complex.
This is nothing but the auxiliary sign problem.

Necessity of the Wick rotation of
the $\omega_0$ field in the usual
NJL model formulation can be seen from the detailed procedure of the
bosonization.
The auxiliary field, $\omega_\mu$, is introduced by inserting $1$ to the
partition function via the Gauss integral to eliminate the four-fermi
interactions;
\begin{align}
 1 &= \int_{C_\mathbb{R}} {\cal D} \omega_0
      \exp \left[ G_\mathrm{v} \int d^4x_E~
      \Bigl( \omega_0(x) - V_0(x) \Bigr)^2 \right]\ ,
\end{align}
where $V_0(x)={\bar q}(x)\gamma_0 q(x)$.
This identity is not
valid since the sign of the $\omega_0^2$
term is not negative and the integral is not well-defined.
Thus, the identity should be modified as
\begin{align}
 1 &= \int_{C_\mathbb{R}} {\cal D} \omega_0
      \exp \left[ - G_\mathrm{v} \int d^4x_E~
	\left(\omega_4(x)+iV_0(x)\right)^2 \right]
\end{align}
Since the sign of the $\omega_4^2$ term becomes
negative, the identity is manifested.
This is the reason why we need the Wick rotation in the usual NJL model
formulation with the repulsive vector-current interaction.
It should be noted that usual NJL model formulation can be acceptable if
we adapt the constraint condition that $\omega_0$ is just identified as
the quark number density.
In this case, we solve the gap equation for $\omega_0$ and do not
(and cannot) perform the path integral about $\omega_0$.

It should be noted that $2 i G_\mathrm{v} \omega_4$ term in $\mu'$
should be $2 G_v \omega_o$ before the Wick rotation and then the
auxiliary sign problem is absent, but the path integral is not well
defined because the Boltzmann weight, $e^{-\Gamma}$, on the integration
path of $\omega_0$ is not stable with fixed $\sigma$;
\begin{align}
\lim_{\omega_0 \to \pm \infty} {\cal V} = - \infty, ~~~
\lim_{\omega_0 \to \pm \infty} (- \Gamma) = + \infty.
\end{align}
Therefore, the path integral is not well-defined without the Wick
rotation of the $\omega_0$ field.

Throughout this paper, we use the parameter set obtained in
Ref.~\cite{Kashiwa:2006rc},
$m_0 = 5.5$ MeV, $G = 5.498$ GeV$^{-2}$, $G_\mathrm{v}=0.25G$ and
the three-dimensional momentum cutoff, $\Lambda = 631.5$ MeV
, which
reproduce empirical values of the pion mass
and decay constant.

\section{Singularities and prescription}
\label{Sec:NR}
We can understand some properties of the
new integration-path contour and
stability of the path integral without performing numerical
calculations.
Therefore, we firstly summarize the properties here;
\begin{align}
&\Gamma[x] \in \mathbb{R},~~
 \lim_{\omega_0 \to \pm \infty} (-\Gamma[z]) = +\infty,
\nonumber\\
& \hspace{2cm}
\begin{CD}
@VV \mathrm{~Wick~rotation} V
 \end{CD}
\nonumber \\
& \Gamma[x] \in \mathbb{C},~~
\lim_{\omega_4 \to \pm \infty} (-\Gamma[z]) = -\infty,
\nonumber\\
& \hspace{2cm}
\begin{CD}
@VV \mathrm{~On~Lefschetz~thimbles} V
 \end{CD}
\nonumber\\
 & \Gamma[z] \in \mathbb{R},~~
\lim_{\omega_4 \to \pm \infty} (-\Gamma[z]) = -\infty,
\label{Eq:set}
\end{align}
where $\sigma$ is fixed in the first and second lines.
The first line means the path integral on $C_\mathbb{R}$.
The second line is the path integral after the Wick rotation (correct
bosonization) for the $\omega_0$ field
and the third one means the path integral on ${\cal J}$ with $\omega_4$.
In the flowchart (\ref{Eq:set}), we assume that the Lefschetz thimbles
do not end at singular points in the estimation of $\Gamma$ in
Eqs.~(\ref{Eq:set}).
However, the stability of the integration still holds by
Eq.~(\ref{Eq:gfc}) even if ${\cal J}$ ends at singular points.

In Eq.~(\ref{Eq:EP-NJL}), there are the square root
and logarithmic functions in the momentum integration.
On the original path $C_\mathbb{R}$, such term does not induce the
difficulty, but these cause the problem in the complexified system.
Since these are multivalued functions, we should care the singular points
and cuts to obtain correct results.
Actually, gradient flows starting from fixed points sometimes show
numerically singular behavior and then we can not continuously draw the
flow trajectories.

In order to discuss the problem coming from the singularities and cuts,
we first consider only the $\sigma$ field, and $\omega_4$ is fixed to $0$.
The thermal part of the effective potential for the particle
contribution ($\Gamma_\mathrm{T}$) becomes
\begin{align}
\Gamma_\mathrm{T} =& \frac{N_cN_f}{3\pi^2}\int_0^\Lambda dp \Bigl[
 \frac{p^4}{\sqrt{p^2+M^2}} \frac{1}{1+e^{(\sqrt{p^2+M^2}-\mu)/T}}
 \Bigr] \nonumber \\
 &- \frac{N_cN_f}{3\pi^2}\Lambda^3T\log(1 +
 e^{\frac{\sqrt{\Lambda^2+M^2}-\mu}{T}})
\end{align}
where we use the integration by parts to remove the logarithm.
We can easily find singular points and cuts induced by $\sigma$ and
$p$-integration as
\begin{align}
z &= i (2k+1) \pi T,
\end{align}
where $z = \sqrt{p^2 + M^2} - \mu$ and $k \in \mathbb{Z}$.

\begin{figure}[t]%[H]
\begin{center}
 \includegraphics[width=0.48\textwidth]{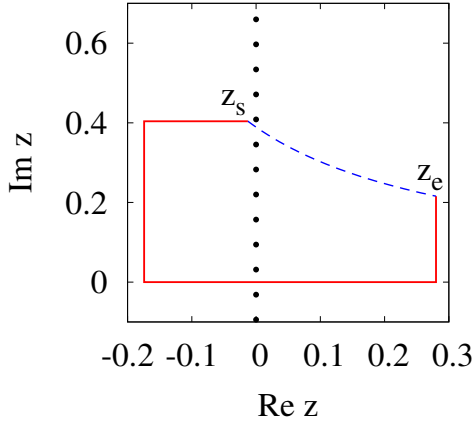}
\end{center}
 \caption{
 The $z$ integration path when the start and end points
 ($z_\mathrm{s}$ and $z_\mathrm{e}$) appear in
 the different quadrants.
 Closed circles represent singular points.
 The solid and dashed lines mean the integration path with and without
 our prescription, respectively.
 }
\label{Fig:path1}
\end{figure}

\begin{figure}[t]%[H]
\begin{center}
\includegraphics[width=0.48\textwidth]{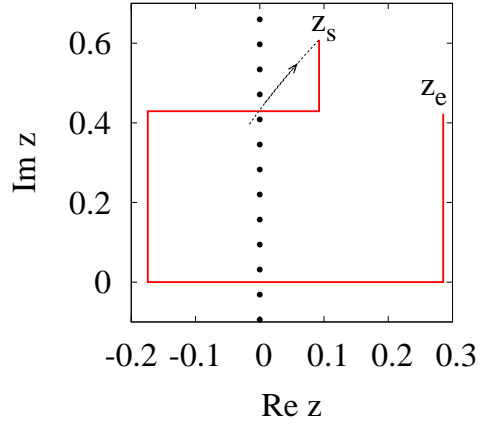}
\end{center}
 \caption{
 The $z$ integration path when $z_\mathrm{s}$ and $z_\mathrm{e}$ appear
 in the same quadrant.
 The arrow means the moving direction of the start position from
 Fig.~\ref{Fig:path1} to present figure with the $t$-evolution of the
 gradient flow.
 Other symbols and lines are same with those in Fig.~\ref{Fig:path1}.
 }
\label{Fig:path2}
\end{figure}

It is natural that gradient flows run on the same Riemann sheet and thus
we extend $p \to {\tilde p} \in \mathbb{C}$ to take care of singularities;
the ${\tilde p}$ integration-path should not go across singular points
and cuts with varying $M$.
Following flowchart shows the prescription of the ${\tilde p}$-modification.
For the convenience of explanations, we use the complex $z$ instead of
${\tilde p}$.
\begin{enumerate}
  \item   Start the calculation with the condition that the start and
	  end points of the $z$-integration, $z_\mathrm{s}$ and
	  $z_\mathrm{e}$, exist in the different quadrant.
  \item   Set the integration path as shown in Fig.~\ref{Fig:path1} to
	  go across the origin.
  \item   Store the positions of $z_\mathrm{s}$ and $z_\mathrm{e}$.
  \item   Evaluate next $z_\mathrm{s}$ and $z_\mathrm{e}$ with the
	  $t$-evolution of gradient flows in Eq.~(\ref{Eq:gf}).
  \item   Repeat the procedure 2 as long as $z_\mathrm{s}$ and
	  $\mathrm{z}_e$ exist in different quadrants.
	  If $z_\mathrm{s}$ and $z_\mathrm{e}$ appear in the same
	  quadrant, the integration path should be
	  taken as Fig.~\ref{Fig:path2} to go across the same slit
	  between singular points where $z_\mathrm{s}$ or $z_\mathrm{e}$
	  go through before.
 \item    Go back to the procedure 3 until the enough length of the
	  Lefschetz thimble is obtained.
 \end{enumerate}
%%%%%%%%%%%%%%%%%%%% Fig %%%%%%%%%%%%%%%%%%%%%%%%
%\begin{figure}[htbp]%[H]
%\begin{figure}[htbp]%[H]
%%%%%%%%%%%%%%%%%%%%%%%%%%%%%%%%%%%%%%%%%%%%%%%%%
This prescription is imposed to fix the flow trajectories on the same
Riemann sheet in the gradient-flow evolution since
the value of the ${\tilde p}$-integration depends on the
slit through which the path goes.
In the naive Lefschetz thimble method without the prescription, the
integration path in the complex $z$ plane becomes the
dashed line in Fig.~\ref{Fig:path1} and it provides wrong results.

%%%%%%%%%%%%%%%%%%%% Fig %%%%%%%%%%%%%%%%%%%%%%%%
%\begin{figure}[htbp]%[H]
\begin{figure}[t]%[H]
\begin{center}
\includegraphics[width=0.48\textwidth]{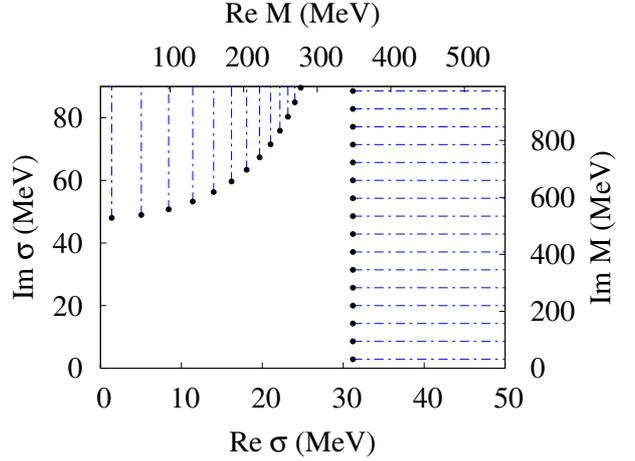}
\end{center}
 \caption{ Structure of singular points and cuts in the complex
 $\sigma$ plane.
 }
\label{Fig:cut}
\end{figure}
%%%%%%%%%%%%%%%%%%%%%%%%%%%%%%%%%%%%%%%%%%%%%%%%%

Figure~\ref{Fig:cut} shows the structure of singular points and cuts
in the complex $\sigma$ plane.
It should be noted that we can set directions of cuts to a certain
degree and thus present directions are just an example.
%%%%%%%%%%%%%%%%%%%% Fig %%%%%%%%%%%%%%%%%%%%%%%%
%\begin{figure}[htbp]%[H]
\begin{figure}[t]%[H]
\begin{center}
\includegraphics[width=0.48\textwidth]{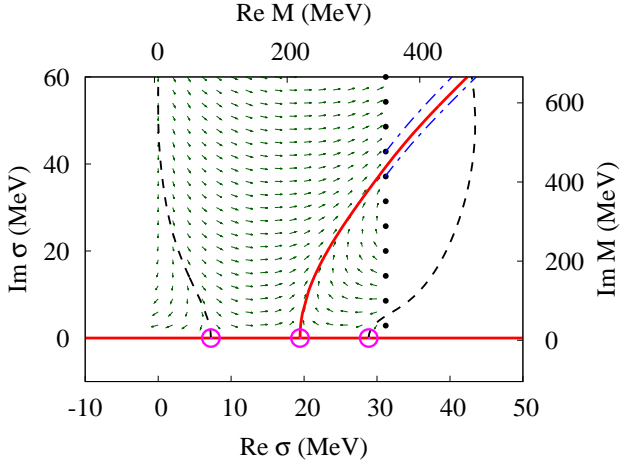}
\end{center}
 \caption{
 Lefschetz thimbles in the complex $\sigma$ plane.
 Open circles mean maximum and minima of the effective
 action, corresponding to $\tau=1,3,2$ respectively from the left.
 Solid and dashed lines represent ${\cal J}$ and
 ${\cal K}$ for each thimble, respectively.
 In the present case, ${\cal J}_{1,2}$ and ${\cal K}_3$ form same
 lines and thus we do no show ${\cal K}_3$ in the figure.
 Singular points are expressed by closed circles.
 Cuts are drawn
 by dot-dashed lines, but we here only draw nearest-neighbor cuts of
 ${\cal J}_2$.
 }
\label{Fig:Fig:thimble-NJL}
\end{figure}
%%%%%%%%%%%%%%%%%%%%%%%%%%%%%%%%%%%%%%%%%%%%%%%%%
The NJL model is the cutoff theory and thus the $|M|>\Lambda$ region
does not matter, but some cuts exist inside the relevant
region for the integration.
This problem should be important in all $\mu$ region.
%To care the singularities and cuts, we should use following
%numerical integration procedures:
%\begin{enumerate}
% \item Check the starting and end points of integration in $z$ plane as
%       Fig.~\ref{Fig:cut1}.
% \item Consider the analytic continuation from the original $z$
%       integration-path to the complex $z$ plane.
% \item Store the trajectory of the starting and end points on the complex
%       $z$ plane with the evolution of variables of integration.
% \item If starting and end points exist in same quadrant,
%       modify the $z$ integration integration path as
%       Fig.~\ref{Fig:cut3}.
% \item
%\end{enumerate}
Results are shown in Fig.~\ref{Fig:Fig:thimble-NJL}.
Present NJL model in the numerical calculation does
not have the repulsive
vector-current interaction and thus it does not have the auxiliary sign
problem.
We just use this model as an laboratory to check the singularity and cut
effects for gradient flows.
Actually, Lefschetz thimbles and the original integration-path become
exactly the same in this case.
Of course, it will be changed if we introduce the vector-current
interaction.

In the present analysis, we analytically sum over the Matsubara
frequencies and thus we encounter the difficulty of singular points and
cuts.
However, it is also true that we will encounter the same difficulty even
in the case that Matsubara frequencies are not summed over analytically
in the the coordinate-space representation because both path-integral
formulations in the momentum- and coordinate-space
representations are equivalent.
In the lattice simulation, it is impossible to take the exact
thermodynamic limit without the extrapolation and thus present
singularity issue may be weaken, but it exists in principle.
Therefore, numerical calculations are safe
if configurations are localized far from the singularities, but not
if configurations appear around the singularities.
This problem should also appear in the complex Langevin method since the
method also use the flow evaluation for performing the integration process.
The difficulty from the singularities becomes more serious when the
dimension of the space-time becomes larger and larger.
Therefore, our analysis becomes important if we apply
the Lefschetz-thimble path-integral method and the complex Langevin
method to the four dimensional and also higher dimensional fermionic models.

\section{Lefschetz thimble of Wick rotated vector field}
\label{Sec:omega}

We shall now discuss the Lefschetz thimble of the Wick rotated vector field.
In the previous section, we have found that the thimble of the auxiliary scalar
field is the same as the original integration path
when the vector field is fixed to be zero.
Then we do not have the sign problem.
In the case where the vector field is switched on,
the auxiliary sign problem may arise as discussed in the Introduction.
We adopt the same prescription as in the scalar field case;
the gradient flow trajectory is required to evolve on the same Riemann sheet.

%%%%%%%%%%%%%%%%%%%% Fig %%%%%%%%%%%%%%%%%%%%%%%%
\begin{figure}[tbhp]%[H]
\begin{center}
\includegraphics[width=0.48\textwidth]{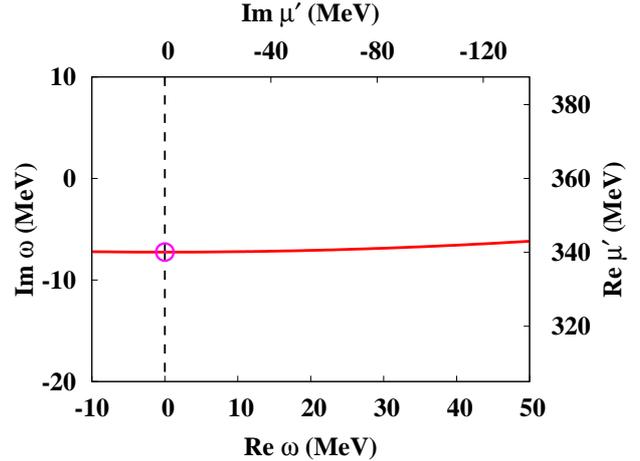}
\end{center}
 \caption{
 Lefschetz thimble in the complex $\omega_4$ plane.
 The solid and dashed lines show the downward and upward thimbles,
${\cal J}$ and ${\cal K}$, respectively,
and the open circle shows the fixed point.
 }
\label{Fig:omega}
\end{figure}
%%%%%%%%%%%%%%%%%%%%%%%%%%%%%%%%%%%%%%%%%%%%%%%%%

In Fig.~\ref{Fig:omega}, we show the Lefschetz thimble for the Wick rotated
auxiliary vector field, $\omega$, at $T=10$ MeV and $\mu=360$ MeV.
The scalar field, $\sigma,$ is small at this $(T,\mu)$ and is
assumed to be zero.
As expected, the $\omega_4$ takes the pure imaginary value
at the fixed point,
$\omega_4=-i\omega_0 \simeq -i \langle q^\dagger q\rangle$,
and the downward thimble runs approximately in
parallel to the original integration
path, the real axis.
Thus, by constructing the integration path of the Wick rotated vector field
in the complex plane,
the auxiliary sign problem can be removed.

\section{Summary}
\label{Sec:Summary}
In this study, we have investigated the auxiliary sign problem which
arises when the fermionic theory has the repulsive vector-field in
variables of integration after the bosonization procedure:
The Boltzmann weight in the partition function should
oscillate by the repulsive vector-field when we make the
path-integral well-define.
If fermionic effective models do not have
the sign problem and those path-integral are ill-defined after the simple
bosonization procedure, the Wick rotation of the repulsive vector-field
cures the illness.
Necessity of this Wick rotation has been discussed in the detailed
procedure of the bosonization.
However, the Wick rotation induces the auxiliary sign problem.
To explicitly discuss the auxiliary sign problem, we have used the
two-flavor Nambu--Jona-Lasinio (NJL) model as an example.
This model with the vector-current interaction dose not have
the original sign problem, but its path-integral is ill-defined.
The Wick rotation for the $\omega_0$ field which is induced from the
repulsive vector-current interaction after the bosonization can make the
path integral well-defined.
However, the $\omega_4$ field induces the oscillating Boltzmann weight
in the NJL partition function via the effective complex chemical
potential.
This auxiliary sign problem can be resolved by the
Lefschetz-thimble path-integral method and then the effective action
satisfy following relations;
\begin{align}
\Gamma[z] \in \mathbb{R},~~~~
\lim_{\omega_4 \to \pm \infty} (-\Gamma[z]) = -\infty.
\end{align}
Therefore, the path integral on the Lefschetz thimbles with the Wick
rotation is well-defined and then the auxiliary sign problem vanishes.

However, after using the Wick rotation and the Lefschetz thimble method,
we found that there are singular points and cuts
induced by the square root and the logarithmic function in the momentum
integration plane.
These singularities sometimes induce the numerical instability for
the gradient-flow evolution and we can not draw the Lefschetz
thimble continuously.
To analyze this problem, we consider the system which has either
auxiliary scalar or vector fields.
Then, we have proposed the new prescription for it by using the momentum
integration in the complex domain to fix the gradient flow trajectories
on the same Riemann sheet.
In the prescription, choice of the integration path in the complexified
momentum space is crucial.

This study is related with the sign problem and also the robustly
defined path-integral formulation when the repulsive vector-field exists
in variables of integration.
The vector field is not a special concept in the path integral and thus
present formulation should have wide application range in several
quantum systems.

The authors thank Yoshimasa Hidaka and Yuya Nakagawa for helpful comments.
A.O. is supported in part by the Grants-in-Aid for Scientific Research
 from JSPS (Nos. 15K05079, 15H03663, 16K05350),
the Grants-in-Aid for Scientific
 Research on Innovative Areas from MEXT (Nos. 24105001, 24105008),
 and by the Yukawa International Program for Quark-hadron
 Sciences (YIPQS).
 
\bibliographystyle{elsarticle-num}
\bibliography{ref}

\end{document}